\documentclass[aps,pre,10pt,twocolumn,nolongbibliography]{revtex4-2}%
\pdfoutput=1
\usepackage{graphicx}%
\usepackage{amsmath, amssymb}%
\usepackage{mathphyscmd}%
\everymath{\displaystyle}%
\usepackage{hyperref}%
\hypersetup{%
	pdfinfo={%
		Author       = {David Gaspard and Jean-Marc Sparenberg},
		Title        = {Resonance distribution in the quantum random Lorentz gas},
		Subject      = {Scattering theory},
		CreationDate = {D:20210917115708},%
	},
	pdfborderstyle={/S/U/W 1},%
	breaklinks=true
}

\begin{document}%
\title{Resonance distribution in the quantum random Lorentz gas}%

\author{David Gaspard}
\email[E-mail:~]{dgaspard@ulb.ac.be}
\author{Jean-Marc Sparenberg}
\affiliation{Nuclear Physics and Quantum Physics, CP229, Universit\'e libre de Bruxelles (ULB), École polytechnique, B-1050 Brussels, Belgium}
\date{\today}

\begin{abstract}%
The multiple scattering model of a quantum particle in a random Lorentz gas consisting of fixed point scatterers is considered in arbitrary dimension.
An efficient method is developed to numerically compute the map of the density of scattering resonances in the complex plane of the wavenumber without finding them one by one.
The method is applied to two collision models for the individual scatterers, namely a resonant model, and a non-resonant hard-sphere model.
The results obtained with the former are compared to the literature.
In particular, the spiral arms surrounding the single-scatterer resonance are identified as proximity resonances.
Moreover, the hard-sphere model is used to reveal previously unknown structures in the resonance density.
Finally, it is shown how Anderson localization affects the distribution of resonance widths, especially in the one-dimensional case.
\end{abstract}%
\maketitle

\section{Introduction}\label{sec:introduction}
The propagation of waves in a disordered ensemble of small scatterers is a widely studied topic in physics.
Among the fundamental tools, a remarkable one is the multiple scattering method which consists in solving the self-consistent equations based on the free-space Green function.
This method, which we develop in the companion paper in the case of a matter-wave quantum particle~\cite{GaspardD2022a}, is commonly used in many fields of application including optics, underwater acoustics~\cite{Lanoy2015, Feuillade1995, Raveau2015, Wout2021}, and electronic band computations in solid state physics~\cite{Korringa1947, Gonis2000}.
This method enjoys great freedom in the choice of the collisional and geometrical properties for the scatterers, such as their positions.
Another advantage is the relatively weak influence of the number of spatial dimensions in the formulation of the problem, as it is also valid in one dimension.
The main limitation is the number of scatterers which can hardly exceed a few tens of thousands even with powerful computing resources~\cite{Lanoy2015}.
\par An additional tool to study the propagation of a quantum particle is the resonance spectrum which extends in the complex plane of the frequency or the wavenumber.
The resonance spectrum provides the knowledge of the time scales controlling the propagation of waves in the medium of interest.
In particular, the imaginary parts of the resonances, i.e., the resonance widths, are inversely proportional to the decay time of the wave within the medium, and so provide useful information about the dynamical processes taking place in the system.
A notable example of a phenomenon which can be evidenced by the distribution of resonance widths is Anderson localization~\cite{Anderson1958, Anderson1978, ShengP2006, Lagendijk2009}.
This phenomenon is characterized by the reduction or disappearance of diffusion of waves in random media, especially when the wavelength is larger than the scattering mean free path.
It is known to play a more significant role in low-dimensional systems and is considered universal in one dimension~\cite{ShengP2006, Ossipov2018}.
\par If the quantum particle propagating in the random medium is only weakly coupled to open channels, then the resonances can be approached by the eigenvalues of an effective non-Hermitian Hamiltonian.
This approach was pioneered by Porter and Thomas in their work on nuclear reactions~\cite{Porter1956, Porter1965}.
In this way, the resonance distribution can be obtained by exploiting the techniques of random matrix theory~\cite{Fyodorov1997a, Fyodorov2015, Kottos2005, Weiss2006, Mitchell2010}.
However, little is known about the resonance distribution in strongly open systems such as the random Lorentz gas model.
\par In the special case of strongly resonant scatterers, it was shown by~\cite{Rusek2000, Pinheiro2004} that certain resonances may be given by the eigenvalues of the Green matrix.
Their approach reveals spiral structures in the resonance distribution which are interpreted as proximity resonances~\cite{HellerEJ1996, Li2003}.
The resonance spectrum was later studied in more details by the authors of Refs.~\cite{Skipetrov2011, Goetschy2011a, Goetschy2011b, Goetschy2013-arxiv, Skipetrov2016b, Skipetrov2018} applying techniques of random matrix theory to the Green matrix.
\par In this exploratory paper, we develop a novel method, that we call the resonance potential method, to calculate the distribution of complex resonances in the general case of an arbitrary collisional model for the individual scatterers.
First, in the special case of resonant scatterers, this method is compared to the eigenvalue method from the literature.
The existence of additional resonances, which are due to non-resonant transport, is brought to light.
Second, it is applied to the non-resonant hard-sphere scatterers of our previous paper~\cite{GaspardD2022a}.
The hard-sphere model is valid over a wider range of wavenumbers than the resonant model, and reveals previously unknown structures in the resonance distribution.
\par This paper is organized as follows.
Section~\ref{sec:multiple-scattering-model} presents the multiple scattering model of point scatterers in arbitrary dimension, as introduced in our previous paper~\cite{GaspardD2022a}.
The resonance potential method used to compute the resonance density is established in Sec.~\ref{sec:potential-function}.
This method is then applied to two typical scattering models for the individual scatterers, namely the resonant model in Sec.~\ref{sec:resonant-model}, and the hard-sphere model in Sec.~\ref{sec:hard-sphere-model}.
The eigenvalue method used in the literature to study the resonant model is discussed in Sec.~\ref{sec:multi-eigenvalues}.
Supplemental calculations regarding the Green function can be found in the~\hyperref[app:green-average]{Appendix}.
\par Throughout this paper, we use the notations $d\in\mathbb{N}_{\geq 1}$ for the number of spatial dimensions, and
\begin{equation}\label{eq:ball-surf-vol}
V_d = \frac{\pi^{\frac{d}{2}}}{\Gamma(\frac{d}{2}+1)} \qquad\text{and}\qquad
S_d = V_d d = \frac{2\pi^{\frac{d}{2}}}{\Gamma(\frac{d}{2})}  \:,
\end{equation}
respectively for the volume and the surface area of the ball of unit radius in $\mathbb{R}^d$.
In Eq.~\eqref{eq:ball-surf-vol}, $\Gamma(z)$ denotes the Gamma function~\cite{Olver2010}.

\section{Multiple scattering model}\label{sec:multiple-scattering-model}
We consider a scalar quantum particle of mass $m$ propagating in a Lorentz gas made of $N$ point scatterers pinned to the random positions $\vect{x}_i$ for $i\in\{1,\ldots,N\}$.
We assume that the positions of the scatterers are contained in a $d$-ball of radius $R$.
The radius $R$ is chosen so as to ensure a uniform density for the gas
\begin{equation}\label{eq:gas-density}
\frac{N}{V_dR^d} = \frac{1}{\varsigma^d}  \:.
\end{equation}
In Eq.~\eqref{eq:gas-density}, $\varsigma$ (sigma) is the mean inter-scatterer distance that we treat as the unit length of the problem.
The stationary wave function $\psi(\vect{r})$ of the quantum particle obeys the Schrödinger equation
\begin{equation}\label{eq:main-schrodinger}
(\nabla^2 + k^2 - U(\vect{r}))\psi(\vect{r}) = 0  \:,
\end{equation}
where $\nabla^2$ is the Laplace operator in $\mathbb{R}^d$, and $k=2\pi/\lambda$ is the wavenumber in free space.
The potential $U(\vect{r})$ in Eq.~\eqref{eq:main-schrodinger} is defined as the sum
\begin{equation}\label{eq:general-multi-potential}
U(\vect{r}) = \sum_{i=1}^N u(\vect{r}-\vect{x}_i)  \:,
\end{equation}
where $u(\vect{r})$ denotes the potential associated with a single scatterer, and the sum runs over the random positions of the scatterers.
Due to the point-like nature of the scatterers, the spatial range of $u(\vect{r})$ is neglected in front of all other characteristic lengths of the problem, especially the wavelength $\lambda$ and the mean inter-scatterer distance~$\varsigma$.
An important consequence of this assumption is that the collisions with the scatterers involve only $s$ waves.

\subsection{Single scatterer}\label{eq:single-scatterer}
In order to solve Eq.~\eqref{eq:main-schrodinger}, one efficient way is to use the Green function method~\cite{Joachain1979, Newton1982, Taylor2006, Gonis2000, ShengP2006, Akkermans2007, Born2019, Rossum1999}.
The Green function, denoted as $G(k,r)$, is defined as a solution of the Schrödinger equation with a point source term at the origin ($\vect{r}=\vect{0}$)
\begin{equation}\label{eq:def-free-green}
(\nabla^2 + k^2)G(k,r) = \delta^{(d)}(\vect{r})  \:.
\end{equation}
Two linearly independent solutions can be found out of Eq.~\eqref{eq:def-free-green}. In arbitrary dimension $d$, they read
\begin{equation}\label{eq:free-green-hankel}
G^\pm(k,r) = \pm\frac{1}{4\I} \left(\frac{k}{2\pi r}\right)^\frac{d-2}{2} H^\pm_{\frac{d-2}{2}}(kr)  \:.
\end{equation}
The function $G^+(k,r)$ is known as the outgoing Green function, and $G^-(k,r)$ is the incoming Green function~\cite{Joachain1979}.
The behavior of these functions is better highlighted by the asymptotic approximation
\begin{equation}\label{eq:free-green-asym}
G^\pm(k,r) \xrightarrow{r\rightarrow\infty} \pm\frac{1}{2\I k}\left(\frac{\mp\I k}{2\pi r}\right)^{\frac{d-1}{2}} \E^{\pm\I kr}  \:.
\end{equation}
Regarding the complex plane of $k$, the two Green functions $G^\pm(k,r)$ possess a branch cut on the negative real axis of $k$ in even dimensions $d\in\{2,4,\ldots\}$.
It turns out that this branch cut does not satisfy the symmetry properties
\begin{equation}\label{eq:green-symmetries}\begin{cases}
\cc{G^\pm(-\cc{k},r)} = G^\pm(k,r)  \:,\\
G^\pm(-k,r) = G^\mp(k,r)  \:,
\end{cases}\end{equation}
which are generally expected for scattering observables~\cite{Joachain1979, Newton1982}.
In order to restore the symmetry properties~\eqref{eq:green-symmetries}, we choose to move this branch cut on the imaginary axis of $k$.
This can be done with the modified Bessel function $K_\nu(z)$~\cite{Olver2010}. We have
\begin{equation}\label{eq:free-green-bessel-k}
G^\pm(k,r) = -\frac{1}{2\pi}\left(\frac{\mp\I k}{2\pi r}\right)^{\frac{d-2}{2}} K_{\frac{d-2}{2}}(\mp\I kr)  \:.
\end{equation}
In this paper, since we deal with arbitrary dimensions, including even ones, we prefer to define the Green functions as in Eq.~\eqref{eq:free-green-bessel-k}, for convenience.
So, the possible branch cut of $G^+(k,r)$ lies on the negative imaginary axis ($\arg k=-\tfrac{\pi}{2}$), and the branch cut of $G^-(k,r)$ on the positive imaginary axis ($\arg k=+\tfrac{\pi}{2}$).
We will come back to this important aspect in Secs.~\ref{sec:potential-function} and~\ref{sec:resonance-density-hard-sphere}.
\par Another useful function related to the Green functions is their imaginary part
\begin{equation}\label{eq:def-free-green-imag}
I(k,r) = -\Im[G^+(k,r)] = -\frac{G^+(k,r) - G^-(k,r)}{2\I} \:.
\end{equation}
This function comes up in many expressions, in particular those concerning the scattering amplitude and the probability conservation law~\cite{GaspardD2022a}.
Note that, in contrast to $G^+(k,r)$, the function $I(k,r)$ behaves as a constant in the neighborhood of the point $r=0$.
At the point $r=0$ itself, it reduces to
\begin{equation}\label{eq:free-green-imag-zero}
I(k,0) = \frac{\pi}{2}\frac{S_d}{(2\pi)^d}k^{d-2}  \:.
\end{equation}
\par Using the Green function~\eqref{eq:free-green-bessel-k}, we can solve Eq.~\eqref{eq:main-schrodinger} in the special case of a single scatterer ($N=1$).
If we assume that the particle collides with the scatterer in the plane-wave state $\phi(\vect{r})=\E^{\I k\vect{\Omega}_0\cdot\vect{r}}$ of initial direction $\vect{\Omega}_0$, we can write
\begin{equation}\label{eq:scat-wave-function-asym}
\psi(\vect{r}) = \phi(\vect{r}) + F(k) G^+(k,\norm{\vect{r}})  \:.
\end{equation}
In Eq.~\eqref{eq:scat-wave-function-asym}, $F(k)$ denotes the scattering amplitude of the scatterer. It is related to the $s$-wave phase shift $\delta(k)$ by
\begin{equation}\label{eq:point-ampli-from-phase-shift}
F(k)^{-1} = I(k,0)\left(\I - \cot\delta(k)\right)  \:.
\end{equation}
As long as the phase shift value is real, a scattering amplitude of the form~\eqref{eq:point-ampli-from-phase-shift} ensures probability conservation during the collision.
This conservation condition can also be expressed as
\begin{equation}\label{eq:unitary-condition}
\Im[F(k)^{-1}] = I(k,0)  \qquad\forall k\in\mathbb{R}  \:.
\end{equation}
The total cross section of the scatterer is related to $F(k)$ by
\begin{equation}\label{eq:total-cross-section}
\sigma_{\rm pt}(k) = \frac{1}{k}I(k,0) \abs{F(k)}^2   \:.
\end{equation}
The conservation condition~\eqref{eq:unitary-condition} requires the total $s$-wave cross section to be smaller than the upper bound~\cite{GaspardD2022a}
\begin{equation}\label{eq:max-cross-section}
\sigma_{\max}(k) = \frac{1}{k\,I(k,0)}  \:.
\end{equation}
\par In this paper, we will consider two different models for $F(k)$.
We will first consider a \emph{resonant model} for the single scatterer described by a Breit-Wigner profile~\cite{Nussenzveig1972, HellerEJ1996, Rusek2000, DeVries1998}.
In the vicinity of the resonance pole $p=\freal{p}+\I\fimag{p}$, this model reads in our notations
\begin{equation}\label{eq:point-ampli-resonant}
F_{\rm rs}(k)^{-1} = I(k,0)\left(\I - \frac{k - \freal{p}}{\fimag{p}}\right)  \:.
\end{equation}
It is assumed that the resonance pole is located in the lower half plane ($\fimag{p}<0$).
The second model that we will consider is the \emph{hard-sphere model} for $s$-wave scattering derived in our previous paper~\cite{GaspardD2022a}
\begin{equation}\label{eq:point-ampli-hard-sphere}
F_{\rm hs}(k)^{-1} = -I(k,0)\frac{G^+(k,\alpha)}{I(k,\alpha)}  \:,
\end{equation}
where $\alpha$ is the scattering length.
The parameter $\alpha$ can be interpreted as the radius of a hard sphere, hence its name, but it has a more general meaning according to scattering theory since it may also be negative~\cite{Newton1982, Joachain1979}.
It turns out that this parameter describes the universal behavior of the $s$-wave scattering amplitude when the scatterer is much smaller than the wavelength~\cite{Bolle1984b, Verhaar1985, GaspardD2018a, GaspardD2022a}.
In this regard, the domain of physical validity of Eq.~\eqref{eq:point-ampli-hard-sphere} is thus $\abs{k}\lesssim\alpha^{-1}$.
\begin{figure}[ht]%
\includegraphics{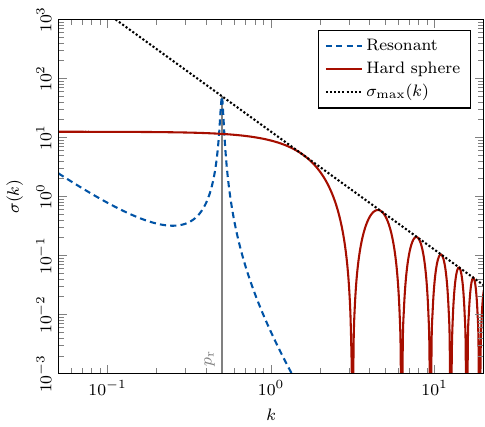}%
\caption{Total cross section of a single scatterer in 3D for the resonant model of Eq.~\eqref{eq:point-ampli-resonant} with $p=0.5-0.01\,\I$, and the hard-sphere model of Eq.~\eqref{eq:point-ampli-hard-sphere} with $\alpha=1$. The gray vertical line highlights the position $\freal{p}$ of the resonance.
The dashed line depicts the maximum cross section of Eq.~\eqref{eq:max-cross-section}.}%
\label{fig:scattering-models}%
\end{figure}%
\par The total cross sections associated with both models~\eqref{eq:point-ampli-resonant} and~\eqref{eq:point-ampli-hard-sphere} are graphically compared in Fig.~\ref{fig:scattering-models}.
In contrast to the resonant model, the hard-sphere model does not display any resonance pole for $k\in\mathbb{C}$.
The oscillations of the hard-sphere model observed in Fig.~\ref{fig:scattering-models} for $\alpha k>1$ are due to non-physical cancellations of the scattering amplitude $F_{\rm hs}(k)$, inherent to this model~\cite{GaspardD2022a}.
On the other hand, only the hard-sphere model exhibits the expected low-energy behavior for $\abs{k}\lesssim\alpha^{-1}$.
Since $\alpha$ can be made arbitrarily small, this domain of validity thus extends in a wider range of wavenumber than Eq.~\eqref{eq:point-ampli-resonant} which is limited to the neighborhood $\abs{k-\freal{p}}\lesssim\abs{\fimag{p}}$ for $\abs{\fimag{p}}\ll\abs{\freal{p}}$.
In short, model~\eqref{eq:point-ampli-resonant} has one resonance but not the expected low-energy behavior, and conversely model~\eqref{eq:point-ampli-hard-sphere} has no resonance but the expected low-energy behavior.
In this regard, both models will provide complementary view points on the resonance distribution.

\subsection{Many scatterers}\label{eq:many-scatterers}
When there are more than one scatterer, the particle wave function can be expressed outside the scatterers as
\begin{equation}\label{eq:multi-wave-function}
\psi(\vect{r}) = \phi(\vect{r}) + \sum_{i=1}^N a_i G^+(k,\norm{\vect{r}-\vect{x}_i})  \:,
\end{equation}
where $a_i$ denotes the complex scattered amplitude one the $i^{\textrm{th}}$ scattering site~\cite{GaspardD2022a}.
These amplitudes satisfy the self-consistent Lippmann-Schwinger equation describing the scattering between all the scatterers.
This equation reads
\begin{equation}\label{eq:lippmann-schwinger}
a_i = F(k)\left(\phi(\vect{x}_i) + \sum_{j (\neq i)}^N a_j G^+(k,r_{ij})\right)  \:,
\end{equation}
where $r_{ij}=\norm{\vect{x}_i-\vect{x}_j}$ is the distance between all pairs of scatterers.
The amplitudes $a_i$ are thus the solution of the linear system 
\begin{equation}\label{eq:m-matrix-lippmann-schwinger}
\matr{M}(k)\,\vect{a} = \vect{\phi}  \:,
\end{equation}
where $\vect{a}=\tran{(a_1,\ldots,a_N)}$ and $\vect{\phi}=\tran{(\phi(\vect{x}_1),\ldots,\phi(\vect{x}_N))}$.
Therefore, the multiple scattering problem is completely described by the matrix $\matr{M}(k)$ defined as
\begin{equation}\label{eq:def-m-matrix}
M_{ij}(k) = F(k)^{-1}\delta_{ij} - G^+(k,r_{ij})(1-\delta_{ij})  \:.
\end{equation}
We refer to this matrix as the \emph{multiple scattering matrix}.
In scattering theory, it may be interpreted as the inverse of the transition matrix~\cite{Gonis2000, GaspardD2022a}.
It is worth noting that $\matr{M}(k)$ is a complex symmetric matrix, but is not Hermitian for $k\in\mathbb{R}$.
This matrix-based method is sometimes known as the Foldy-Lax method~\cite{Foldy1945, Lax1951}, or, in the framework of solid-state physics when higher-order partial waves are included, as the Kohn-Korringa-Rostoker method~\cite{Korringa1947, Gonis2000}.
\par Furthermore, according to Eq.~\eqref{eq:m-matrix-lippmann-schwinger}, the scattered amplitudes read $\vect{a}=\matr{M}(k)^{-1}\vect{\phi}$.
This expression predicts the existence of infinite-norm solutions for $\vect{a}$ at the singular values of $k$ given by the determinantal equation
\begin{equation}\label{eq:determinantal-equation}
\det\matr{M}(k) = 0  \:.
\end{equation}
This equation for $k\in\mathbb{C}$ also gives the poles of the transition matrix, and is the central equation of this paper.
Another way of understanding Eq.~\eqref{eq:determinantal-equation} is to look for the non-trivial solutions of Eq.~\eqref{eq:m-matrix-lippmann-schwinger} which exist in the absence of incident wave ($\vect{\phi}=\vect{0}$).
Indeed, the linear system~\eqref{eq:m-matrix-lippmann-schwinger} is then homogeneous and takes the form of a nonlinear eigensystem.
It is nonlinear because $\matr{M}(k)$ does not depend linearly on the ``eigenvalue'', a role played here by the complex wavenumber~$k$.
Once the value of $k$ canceling the determinant of $\matr{M}(k)$ is found, the corresponding eigenvector, $\vect{a}$, can be physically interpreted as a meaningful state.
\par The solutions of Eq.~\eqref{eq:determinantal-equation} may be split in two categories and interpreted differently depending on their location in the complex plane of $k$.
The solutions with negative imaginary part ($\fimag{k}<0$) are interpreted as resonances, and the solutions one the positive imaginary semi-axis ($\arg k=\tfrac{\pi}{2}$) are interpreted as eigenstates~\cite{Joachain1979, Newton1982}.
In addition, the imaginary part of the resonances given by Eq.~\eqref{eq:determinantal-equation} is directly related to the temporal properties.
Indeed, if we consider a complex resonance located at $k=\freal{k}+\I\fimag{k}$ with $\fimag{k}<0$, the imaginary part of the frequency $\omega(k)$ given by the dispersion relation reads
\begin{equation}\label{eq:resonant-state-frequency}
\Im[\omega(\freal{k}+\I\fimag{k})] = v(\freal{k})\fimag{k} + \bigo(\fimag{k}^3)  \:,
\end{equation}
where $v=\partial\omega/\partial k$ is the group velocity of the wave~\cite{Born2019}.
Equation~\eqref{eq:resonant-state-frequency} assumes a relatively small value of $\fimag{k}$ in general, but is exact for both linear or quadratic dispersion relations.
The square modulus of the wave function thus behaves in time as
\begin{equation}\label{eq:resonant-state-evolution}
\abs{\psi(t)}^2 \propto \abs{\E^{-\I\omega(k)t}}^2 = \E^{2\Im[\omega(k)]t} = \E^{2v\fimag{k}t} = \E^{-\Gamma t}  \:.
\end{equation}
In Eq.~\eqref{eq:resonant-state-evolution}, one can identify the quantity
\begin{equation}\label{eq:escape-rate}
\Gamma = 2v(\freal{k})\abs{\fimag{k}}  \:,
\end{equation}
as the characteristic escape rate of the particle from the system when starting in the resonant state $k=\freal{k}+\I\fimag{k}$~\cite{Nussenzveig1972}.
Expression~\eqref{eq:escape-rate} shows that the imaginary part of the resonance position is always proportional to the escape rate, whichever the dispersion relation given by $\omega(k)$.

\subsection{Case of two scatterers}\label{sec:two-atoms}
We consider the special case of a system containing only two scatterers separated by a variable distance $s$.
This case gives valuable qualitative information about the influence of the model parameters on the resonance positions, especially the dimension.
If $N=2$, then the multiple scattering matrix~\eqref{eq:def-m-matrix} reads
\begin{equation}\label{eq:two-atoms-m-matrix}
\matr{M}(k) = \left(\begin{array}{cc}F(k)^{-1} & -G^+(k,s)\\ -G^+(k,s) & F(k)^{-1}\end{array}\right)  \:.
\end{equation}
The complex resonances of the problem are given by the values of $k$ which satisfy $\det\matr{M}(k)=0$ or, equivalently, which send at least one eigenvalue of $\matr{M}(k)$ to zero. This leads to the equation for $k\in\mathbb{C}$ 
\begin{equation}\label{eq:two-atoms-resonances-exact}
G^+(k,s) = \pm F(k)^{-1}  \:.
\end{equation}
Albeit unsolvable for general $F(k)$, approximations of certain solutions to Eq.~\eqref{eq:two-atoms-resonances-exact} can nevertheless be obtained for $ks\gg 1$.
Indeed, in this regime, the Green function behaves as
\begin{equation}\label{eq:free-green-envelope}
G^+(k,s) = A(k,s) \E^{\I ks}  \:,
\end{equation}
according to Eq.~\eqref{eq:free-green-asym}.
Expression~\eqref{eq:free-green-envelope} separates the rapid variations of $\E^{\I ks}$ from the slowly varying envelope $A(k,s)$.
In this way, it is possible to isolate $k$ from the exponential.
This operation results in an infinite number of roots for Eq.~\eqref{eq:two-atoms-resonances-exact} given by
\begin{equation}\label{eq:two-atoms-resonances-approx-1}
k_ns = n\pi - \I\ln\!\left(\frac{F(k_n)^{-1}}{A(k_n,s)}\right)  \:,
\end{equation}
for $n\in\mathbb{Z}$.
If, in addition, one assumes that $F(k)$ is neither infinite nor zero in the region of interest, then the second term of Eq.~\eqref{eq:two-atoms-resonances-approx-1} is relatively small compared to $n\pi$.
Note that this assumption excludes the possible presence of a resonance pole for $F(k)$ in this region.
This allows to use the approximation $k_n\simeq n\pi/s$ for large $n$ in the right-hand side of Eq.~\eqref{eq:two-atoms-resonances-approx-1}.
The solutions of Eq.~\eqref{eq:two-atoms-resonances-approx-1} can thus be approximated by
\begin{equation}\label{eq:two-atoms-resonances-approx-2}
k_ns \simeq n\pi - \I\ln\!\left(\frac{F(n\pi/s)^{-1}}{A(n\pi/s,s)}\right)  \:.
\end{equation}
Equation~\eqref{eq:two-atoms-resonances-approx-2} predicts the existence of a quasi-periodic band of resonances for almost any scattering model $F(k)$ under the stated assumptions.
It also applies, for instance, to the resonant model~\eqref{eq:point-ampli-resonant} far away from the resonance pole.
Moreover, when the number of scatterers $N$ increases, this band is expected to fill with extra resonances until it becomes nearly continuous.
This structure is a universal property of the multiple scattering model, as we will see later.
\begin{figure}[ht]%
\includegraphics{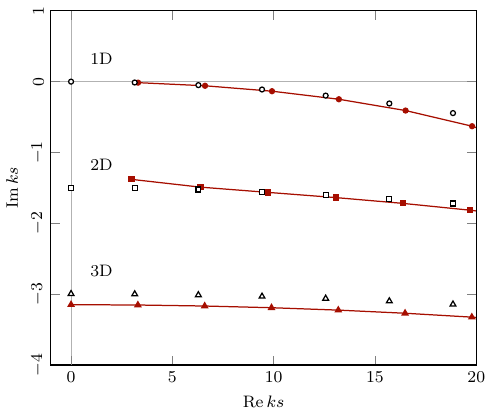}%
\caption{Positions of the complex resonances for two point scatterers in the first three dimensions ($d=1,2,3$) using the hard-sphere $s$-wave scattering model~\eqref{eq:point-ampli-hard-sphere} with $\alpha/s=0.05$.
In red filling, the exact resonances calculated from Eq.~\eqref{eq:two-atoms-resonances-exact}, and, in white filling, the approximations given by Eq.~\eqref{eq:two-atoms-resonances-hard-sphere}.}
\label{fig:resonances-two-atoms}%
\end{figure}%
\par In the special case of the hard-sphere $s$-wave model of Eq.~\eqref{eq:point-ampli-hard-sphere}, additional simplifications can be carried out in Eq.~\eqref{eq:two-atoms-resonances-approx-2}.
Under the assumption that the scattering length is much smaller than the wavelength ($\alpha k\ll 1$), one has
\begin{equation}\label{eq:free-green-imag-series}
\frac{I(k,\alpha)}{I(k,0)} = 1 - \frac{(\alpha k)^2}{2d} + \bigo((\alpha k)^4)  \:.
\end{equation}
Expression~\eqref{eq:two-atoms-resonances-approx-2} then becomes
\begin{equation}\label{eq:two-atoms-resonances-hard-sphere}
k_ns \simeq n\pi - \I\frac{d-1}{2}\ln\!\left(\frac{s}{\alpha}\right) - \frac{\I}{2d}\left(\frac{\alpha}{s}n\pi\right)^2  \:.
\end{equation}
The approximate resonances from Eq.~\eqref{eq:two-atoms-resonances-hard-sphere} are compared to the solutions of Eq.~\eqref{eq:two-atoms-resonances-exact} in Fig.~\ref{fig:resonances-two-atoms}.
The resonances move away from the real $k$ axis as $d$ increases, which means that the particle escapes faster from the system in higher dimensions.
According to Eq.~\eqref{eq:escape-rate} the escape rate corresponding to Eq.~\eqref{eq:two-atoms-resonances-hard-sphere} is approximately given by
\begin{equation}\label{eq:two-atoms-escape-rate}
\Gamma_n\simeq \frac{v}{s}(d-1)\ln\!\left(\frac{s}{\alpha}\right) + \frac{v}{sd}\left(\frac{\alpha}{s}n\pi\right)^2  \:.
\end{equation}
Similar logarithmic behaviors of the escape rate were obtained in Ref.~\cite{GaspardP1989a, *GaspardP1989b} in the context of the quantum scattering on three hard disks.
\par Note that the escape rate~\eqref{eq:two-atoms-escape-rate} decreases with the separation distance between both scatterers.
This may be seen paradoxical because the particle is less tightly enclosed when $s\rightarrow\infty$.
This decrease is due to the factor $\tau=s/v$ in Eq.~\eqref{eq:two-atoms-escape-rate}, which can be interpreted as half the round-trip time of the particle.
As $s\rightarrow\infty$, the round-trip rate falls, hence the decrease of the escape rate $\Gamma_n$.
\par In the one-dimensional case, the very small rate $\Gamma_n\simeq\bigo(v\alpha^2/s^3)$ is due to the fact that the particle cannot escape the corral formed by the two scatterers without crossing them.
This strongly contrasts with the higher dimensional cases for which the periodically bouncing trajectory is classically unstable, leading to much larger escape rates.

\subsection{Resonance potential method}\label{sec:potential-function}
The main concern of this paper is to find the complex scattering resonance poles of the random Lorentz gas for $k\in\mathbb{C}$ satisfying the determinantal equation~\eqref{eq:determinantal-equation}, especially for a large number of scatterers ($N\rightarrow\infty$).
First, it should be noted that one cannot in general solve Eq.~\eqref{eq:determinantal-equation} as a simple eigenvalue problem for $k$ due to the nonlinear dependence of $\matr{M}(k)$ in $k$. This heavily complicates the study of the solutions of this equation.
Despite of this, there are several computational ways to locate the resonance poles in the complex $k$ plane.
One can for instance consider using root-finding methods to locate them one by one.
According to our numerical investigations, it appears that such an algorithm can be noticeably accelerated applying the root-finding method on the smallest eigenvalue of $\matr{M}(k)$ obtained through inverse power iteration, instead of $\det\matr{M}(k)$.
This is because $\det\matr{M}(k)$ behaves exponentially for $k\in\mathbb{C}$, hence dramatically slowing down any Newton-type root-finding method~\cite{Olver2010}.
On the contrary, the smallest eigenvalue of $\matr{M}(k)$ does not vary so much for $k\in\mathbb{C}$, which makes it more suitable for root finding.
\par However, these direct approaches are flawed.
First, the computational time of collecting a significant amount of roots to achieve a statistical analysis can be prohibitive.
Second, given the presumably high density of roots of $\det\matr{M}(k)$, it is hard to ensure that all the roots of a certain region of the complex $k$ plane will have been effectively found by the algorithm.
Indeed, since $\det\matr{M}(k)$ has infinitely many roots for $k\in\mathbb{C}$, starting from initial guesses may bias the observed root distribution due to missing roots.
In this section, we propose a more reliable method to address these issues.
\par Instead of looking for the individual resonances from Eq.~\eqref{eq:determinantal-equation}, it seems more efficient to compute a distribution function.
In this regard, we define the joint resonance density in the complex $k$ plane as
\begin{equation}\label{eq:def-resonance-density}
\rho^{(2)}(k) = \frac{1}{N} \avg{\sum_{j=1}^\infty \delta^{(2)}(k-k_j)}  \:,
\end{equation}
where the sum runs over the roots of $\det\matr{M}(k)$, denoted as $k_j$, and $\delta^{(2)}(k-k_j)=\delta(\Re k-\Re k_j)\delta(\Im k-\Im k_j)$ is a notation for the two-dimensional Dirac delta.
The average of Eq.~\eqref{eq:def-resonance-density} is taken over the random configurations of the scatterer positions.
In order to compute the density~\eqref{eq:def-resonance-density}, we introduce the ancillary function $\Phi:\mathbb{C}\rightarrow\mathbb{R}$ defined as
\begin{equation}\label{eq:def-resonance-potential}
\Phi(k) = \frac{1}{N}\avg{\ln\abs{\det\matr{M}(k)}}  \:,
\end{equation}
that we refer to as the \emph{resonance potential}.
Many equivalent expressions can be written in place of Eq.~\eqref{eq:def-resonance-potential} exploiting the property, $\ln\abs{z}=\Re\ln z$, and the Fredholm-Plemelj formula, $\ln\det\matr{M}=\Tr(\ln\matr{M})$~\cite{Fredholm1903, Plemelj1904, Bornemann2010}, for instance.
\par We can show that the density~\eqref{eq:def-resonance-density} can be derived from Eq.~\eqref{eq:def-resonance-potential}.
First of all, note that this method is restricted to the domain of analyticity of $\det\matr{M}(k)$.
Since the singularities of $\det\matr{M}(k)$ are the same as those of the matrix elements $F(k)^{-1}$ and $G^+(k,r)$, this domain reads
\begin{equation}\label{eq:det-m-analytic-domain}
\mathcal{D} = \begin{cases}%
\{k\in\mathbb{C}\mid F(k)\neq 0 \cap \arg k\neq -\tfrac{\pi}{2}\}  & \text{for even}~d  \:,\\
\{k\in\mathbb{C}\mid F(k)\neq 0\}  & \text{for odd}~d  \:.
\end{cases}\end{equation}
So, $\mathcal{D}$ almost completely encompasses the complex $k$ plane, except for the isolated zeros of $F(k)$ for $k\in\mathbb{R}$, and the branch cut of $G^+(k,r)$ for $\arg k=-\tfrac{\pi}{2}$ in even dimensions.
As a reminder, this branch cut has been placed on the negative imaginary $k$ axis in Eq.~\eqref{eq:free-green-bessel-k} to impose the symmetry of imaginary axis ($k\leftrightarrow-\cc{k}$) that is generally encountered for observables in scattering theory~\cite{GaspardD2022a}.
Furthermore, according to the Weierstrass product theorem~\cite{Olver2010}, $\det\matr{M}(k)$ admits a factorization of the form
\begin{equation}\label{eq:det-m-factorization}
\det\matr{M}(k) = \prod_{j=1}^\infty (k-k_j)\E^{f_j(k)}  \qquad\forall k\in\mathcal{D} \:,
\end{equation}
where $f_j(k)~\forall j$ is some sequence of analytic functions in~$\mathcal{D}$.
Inserting Eq.~\eqref{eq:det-m-factorization} into the resonance potential~\eqref{eq:def-resonance-potential} leads to
\begin{equation}\label{eq:resonance-potential-from-roots}
\Phi(k) = \frac{1}{N} \avg{\sum_{j=1}^\infty f_j(k)} + \frac{1}{N} \avg{\sum_{j=1}^\infty \ln\abs{k-k_j}}  \:.
\end{equation}
Now, we can use the fact that the logarithm is the solution of the two-dimensional Poisson equation
\begin{equation}\label{eq:logarithm-laplacian}
\left(\pder[2]{}{\freal{k}} + \pder[2]{}{\fimag{k}}\right)\ln\abs{k} = 2\pi\delta^{(2)}(k)  \:.
\end{equation}
This implies that the Laplacian of the resonance potential gives the sought resonance density~\eqref{eq:def-resonance-density} up to a factor
\begin{equation}\label{eq:resonance-density-from-potential}
\left(\pder[2]{}{\freal{k}} + \pder[2]{}{\fimag{k}}\right)\Phi(k) = 2\pi\rho^{(2)}(k)  \:.
\end{equation}
Note that the Laplacian of the functions $f_j(k)~\forall j$ in Eq.~\eqref{eq:resonance-potential-from-roots} is equal to zero for $k\in\mathcal{D}$ due to the Cauchy-Riemann equations~\cite{Olver2010} and the fact that the functions $f_j(k)$ are analytic in~$\mathcal{D}$.
\par Note that, according to the Poisson equation~\eqref{eq:resonance-density-from-potential}, $\Phi(k)$ can be geometrically interpreted as a fictitious two-dimensional electrostatic potential generated by point charges located at the complex resonances.
In fact, similar methods are used in random matrix theory to study the distribution of the complex eigenvalues of non-Hermitian matrices~\cite{Goetschy2013-arxiv}.
The difference, here, is that the determinantal equation~\eqref{eq:determinantal-equation} giving rise to the complex resonances is not linear in $k$.
\par Once the resonance potential is computed from Eq.~\eqref{eq:def-resonance-potential} on a rectangular lattice, we numerically evaluate the resonance density from the discrete Laplacian of~$\Phi(k)$.
The advantage of this method is that the mere computation of $\Phi(k)$ from Eq.~\eqref{eq:def-resonance-potential} is much more reliable and faster than finding the roots of $\det\matr{M}(k)$.
This allows us to compute average densities over a large number of random configurations of the scatterer positions.
In addition, this method directly provides the two-dimensional histogram~\eqref{eq:def-resonance-density} without resorting to smoothing kernel techniques.

\section{Resonant scatterers}\label{sec:resonant-model}
In this section, we consider a model in which the scattering amplitude $F(k)$ is given by Eq.~\eqref{eq:point-ampli-resonant} and thus displays a resonance pole at $p=\freal{p}+\I\fimag{p}$ with $\fimag{p}<0$.
This model mostly affects the resonance density of the multiple scattering problem in the vicinity of the single-scatterer resonance at $k=p$.
This special case has been broadly studied in the literature due to the remarkable simplifications that occur near the resonance~\cite{HellerEJ1996, Rusek2000, Pinheiro2004}, as we will see.
This case also gives us the opportunity to compare the resonance potential method, which directly works in the complex $k$ plane (Sec.~\ref{sec:resonance-density-resonant-model}), with the method previously used in the literature~\cite{HellerEJ1996, Rusek2000, Pinheiro2004}, but only indirectly related to the resonances (Sec.~\ref{sec:multi-eigenvalues}).

\subsection{Eigenvalue approach of the resonances}\label{sec:multi-eigenvalues}
The eigenvalue approach, which was originally developed in Refs.~\cite{HellerEJ1996, Rusek2000, Pinheiro2004}, consists of the approximation that, in a small enough neighborhood of $k=p$, the matrix elements of $\matr{M}(k)$ do not strongly vary, except for the diagonal entry $F(k)^{-1}$ which vanishes.
This will be the case if one assumes that the variation of the wavelength in this neighborhood is smaller than the maximum distance between two scatterers, denoted as $L$.
In other words, one has to impose that $\abs{k-p}\ll L^{-1}$.
If, in addition, one assumes that the single-scatterer resonance is narrow ($\abs{\fimag{p}}\ll\abs{\freal{p}}$), then the matrix $\matr{M}(k)$ can be expanded near $k=\freal{p}$ instead of $k=p$. Hence, one finds
\begin{equation}\label{eq:m-pole-approx}
\matr{M}(k) \simeq I(\freal{p},0)\left(\frac{k-\freal{p}}{\fimag{p}} + \matr{N}(\freal{p})\right)  \:,
\end{equation}
where $\matr{N}(\freal{p})$ is a dimensionless matrix defined for convenience as
\begin{equation}\label{eq:def-normalized-matrix}
N_{ij}(k) = \I\,\delta_{ij} - \frac{G^+(k,r_{ij})}{I(k,0)}(1-\delta_{ij})  \:.
\end{equation}
An important point is that $\matr{N}(k)$ does not explicitly depend on the scattering model, in contrast to $\matr{M}(k)$.
Under approximation~\eqref{eq:m-pole-approx}, the determinantal equation~\eqref{eq:determinantal-equation} reduces to a simple eigenvalue problem for $k$.
Denoting the eigenvalues of the matrix $\matr{N}(k)$ as $\nu_j(k)$ for $j\in\{1,\ldots,N\}$, one gets the resonances
\begin{equation}\label{eq:resonance-pole-approx}
k_j \simeq \freal{p} + \fimag{p} \nu_j(\freal{p})  \:.
\end{equation}
Note that a similar approach may be followed to approximate the eigenstates instead of the resonances.
In that case, the pole is purely imaginary ($\freal{p}=0$) and the expansion of Eq.~\eqref{eq:m-pole-approx} should be carried out near $k=\I\fimag{p}$.
Note, in addition, that Eq.~\eqref{eq:resonance-pole-approx} may be interpreted as a generalization of the splitting of degenerate energy levels to resonances.
The role of the perturbation is played by the interaction between the scatterers due to the multiple collisions of the particle.
\par Equation~\eqref{eq:resonance-pole-approx} motivates the interest in the distribution of the complex eigenvalues of $\matr{N}(k)$, especially for real values $k=\freal{p}$.
The spectrum of this matrix has been extensively studied in the three-dimensional case by the authors of Refs.~\cite{Skipetrov2011, Goetschy2011a, Goetschy2011b, Goetschy2013-arxiv} using techniques of random matrix theory.
Here, we recall only the main properties of the eigenvalue distribution of this matrix.
First of all, it is important to note that, the eigenvalues of $\matr{N}(\freal{p})$ have a positive imaginary part
\begin{equation}\label{eq:nu-positive-imag}
\Im\nu_j(\freal{p}) > 0  \:.
\end{equation}
This property can be shown using the positive-definiteness of the quadratic form of the total cross section, assuming real wavenumbers~\cite{GaspardD2022a}.
A consequence of Eq.~\eqref{eq:nu-positive-imag} is that the resonances given by the approximation~\eqref{eq:resonance-pole-approx} have the same sign of their imaginary part as $\fimag{p}$.
Since we have assumed that $\fimag{p}<0$, this means that $\Im k_j<0$, as it should be for resonances~\cite{Joachain1979, Newton1982}.
Moreover, due to the definition~\eqref{eq:def-normalized-matrix}, the average position of the eigenvalues of $\matr{N}(\freal{p})$ is exactly located at
\begin{equation}\label{eq:nu-average-position}
\avg{\nu} = \frac{1}{N}\avg{\Tr\matr{N}(\freal{p})} = \I  \:,
\end{equation}
where $\tavg{\cdot}$ denotes the average over the random positions $\vect{x}_1,\ldots,\vect{x}_N$ of the scatterers.
Therefore, the average position of the resonances given by Eq.~\eqref{eq:resonance-pole-approx} is $\avg{k}=\freal{p}+\I\fimag{p}$, and thus coincides with the single-scatterer resonance $k=p$.
\begin{figure}[ht]%
\includegraphics{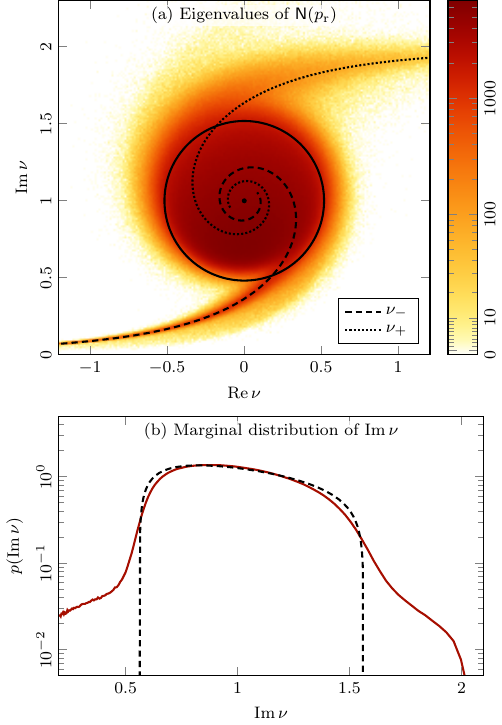}%
\caption{Distribution of the eigenvalues of $\matr{N}(\freal{p})$ in a 3D ball-shaped Lorentz gas for $N=100$ and $\freal{p}=10\,\varsigma^{-1}$, combining $2^{18}$ random configurations.
Panel (a): Full distribution in the complex $\nu$ plane. The spirals are obtained from Eq.~\eqref{eq:nu-eigenvalues-two-atoms} for different separation distances.
The circle in solid line is given by Eqs.~\eqref{eq:nu-average-position} and~\eqref{eq:nu-radius}.
Panel (b): Corresponding marginal distribution $p(\Im\nu)$ (solid), and a fitted Marchenko-Pastur distribution from Eq.~\eqref{eq:marchenko-pastur-law} (dashed).}
\label{fig:mu-plane-d3-n100}%
\end{figure}%
\par Regarding the numerical distribution of the eigenvalues of $\matr{N}(\freal{p})$, two cases are observed depending on the number of scatterers.
First, when the number of scatterers is moderate, the distribution of eigenvalues is shown in Fig.~\ref{fig:mu-plane-d3-n100}.
Several structures are salient: the spiral arms and the circular cluster.
Although Fig.~\ref{fig:mu-plane-d3-n100} is computed for $d=3$, the same structures can be observed in any dimension $d\geq 2$.
The pair of spiral arms in Fig.~\ref{fig:mu-plane-d3-n100}(a) comes from the eigenvalues of the two-scatterer problem.
Indeed, in dimensions $d\geq 2$, the Green function $G^+(\freal{p},r)$ displays a singularity at $r=0$.
So, when two scatterers located at $\vect{x}_i$ and $\vect{x}_j$, for $i\neq j$, get closer to each other, the off-diagonal matrix elements $N_{ij}=N_{ji}$ tend to infinity and dominate the rest of the matrix.
If one neglects the other matrix elements, one gets the two-by-two submatrix
\begin{equation}\label{eq:submatrix-two-atoms}
\matr{N}(\freal{p}) \sim \left(\begin{array}{cc}%
\I & -\frac{G^+(\freal{p},s)}{I(\freal{p},0)}  \\
-\frac{G^+(\freal{p},s)}{I(\freal{p},0)} & \I\end{array}\right)  \:,
\end{equation}
with $s=\norm{\vect{x}_i-\vect{x}_j}$, and thus the approximate eigenvalues
\begin{equation}\label{eq:nu-eigenvalues-two-atoms}
\nu_\pm(\freal{p}) = \I \mp \frac{G^+(\freal{p}, s)}{I(\freal{p}, 0)}  \:.
\end{equation}
These eigenvalues lead to the spiral arms of Fig.~\ref{fig:mu-plane-d3-n100}(a) when the separation distance $s$ varies.
In the limit $s\rightarrow 0$, the spiral arms possess the horizontal asymptotes $\Im\nu_+\rightarrow 2$ and $\Im\nu_-\rightarrow 0$.
The eigenvectors of Eq.~\eqref{eq:submatrix-two-atoms} associated with Eq.~\eqref{eq:nu-eigenvalues-two-atoms} are
\begin{equation}\label{eq:eigenvectors-two-atoms}
\vect{v}_\pm = \frac{1}{\sqrt{2}}\left(\begin{array}{c}1\\ \pm 1\end{array}\right)  \:.
\end{equation}
The state $\vect{v}_+$ is thus symmetric with respect to the permutation of both scatterers, and the state $\vect{v}_-$ is antisymmetric.
So, the wave functions associated with $\vect{v}_+$ and $\vect{v}_-$ resemble to $s$ and $p$ waves, respectively.
These spiral structures were first highlighted in Ref.~\cite{Rusek2000} where they are referred to as \emph{proximity resonances}, a name coined in Ref.~\cite{HellerEJ1996}.
Analogous states are encountered in the context of superradiance~\cite{Akkermans2008, Goetschy2013-arxiv}.
\par Furthermore, the presence of a nearly circular distribution in the center of Fig.~\ref{fig:mu-plane-d3-n100}(a) can be understood as a consequence of the Girko-Ginibre circular law~\cite{Ginibre1965, Tao2008, Gotze2010, Livan2018}.
Indeed, when $\freal{p}$ is large enough to destroy any correlation between the matrix elements of $\matr{N}(\freal{p})$, one can assume that the matrix elements are independent and identically distributed complex random variables and then apply the theory of non-Hermitian ensemble as a first approximation.
According to the circular law, the density of eigenvalues is approximately constant within the cluster and the radius $\rho$ of the distribution is given by~\cite{Goetschy2013-arxiv}
\begin{equation}\label{eq:nu-radius}
\rho^2 = 2\Var(\nu) \simeq \frac{1}{N} \frac{\avg{\Tr[\herm{\matr{G}}(\freal{p})\matr{G}(\freal{p})]}}{I(\freal{p},0)^2}  \:,
\end{equation}
where $\Var(\nu)=\tavg{\abs{\nu}^2}-\abs{\tavg{\nu}}^2$ is the variance of the spectrum, and $\matr{G}(\freal{p})$ is the so-called Green matrix~\cite{Skipetrov2011, Goetschy2011a, Goetschy2011b, Goetschy2013-arxiv} defined as
\begin{equation}\label{eq:def-green-matrix}
G_{ij}(k) = G^+(k,r_{ij})(1 - \delta_{ij}) \:.
\end{equation}
The approximation in Eq.~\eqref{eq:nu-radius} holds for sufficiently large wavenumber ($k\varsigma\gg 1$).
Since all the scatterers are identically and uniformly distributed in the Lorentz gas, the trace in Eq.~\eqref{eq:nu-radius} reduces to
\begin{equation}\label{eq:trace-g2-from-green}
\avg{\Tr[\herm{\matr{G}}(\freal{p})\matr{G}(\freal{p})]} = N(N-1)\avg{\abs{G^+(\freal{p},s)}^2}  \:.
\end{equation}
In Eq.~\eqref{eq:trace-g2-from-green}, $\tavg{\cdot}$ denotes the average over the distance $s=\norm{\vect{x}_i-\vect{x}_j}$ between any pair of scatterers, $\vect{x}_i$ and $\vect{x}_j$, in the gas.
In the three-dimensional case, this average reads for real-valued wavenumbers
\begin{equation}\label{eq:avg-g2-real-d3}
\avg{\abs{G^+(\freal{p},s)}^2} = \left(\frac{3}{8\pi R}\right)^2  \:,
\end{equation}
as shown in the~\hyperref[app:green-average]{Appendix}. More generally, the average~\eqref{eq:avg-g2-real-d3} behaves as $\abs{\freal{p}}^{d-3}/R^{d-1}$.
The circle of center~\eqref{eq:nu-average-position} and radius~\eqref{eq:nu-radius} is compared to the circular cluster in Fig.~\ref{fig:mu-plane-d3-n100}(a).
Note that, under the constraint of unit density ($N=V_dR^d$), the radius behaves as $\rho\propto N^{1/(2d)}$.
So, although the density of the Lorentz gas is kept constant, the circular cluster of eigenvalues still grows with~$N$.
\par When the number of scatterers is limited as in Fig.~\ref{fig:mu-plane-d3-n100}, the marginal distribution of the imaginary parts of $\nu$ can be approximated by the Marchenko-Pastur distribution~\cite{Skipetrov2011, Goetschy2013-arxiv, Livan2018}
\begin{equation}\label{eq:marchenko-pastur-law}
p_{\rm MP}(x) = \frac{\sqrt{(\lambda_+ - x)(x - \lambda_-)}}{Cx}  \:,
\end{equation}
for $x\in[\lambda_-,\lambda_+]$ where $C=\tfrac{\pi}{2}\big(\lambda_+^{1/2} - \lambda_-^{1/2}\big)^2$ is a normalization coefficient.
This well-known distribution gives the eigenvalue density of positive-definite random matrices~\cite{Livan2018}.
Its occurrence in this context is due to the constraint~\eqref{eq:nu-positive-imag} on the eigenvalues.
In Fig.~\ref{fig:mu-plane-d3-n100}(b), the Marchenko-Pastur distribution in dashed is fitted to the numerical data.
\begin{figure}[ht]%
\includegraphics{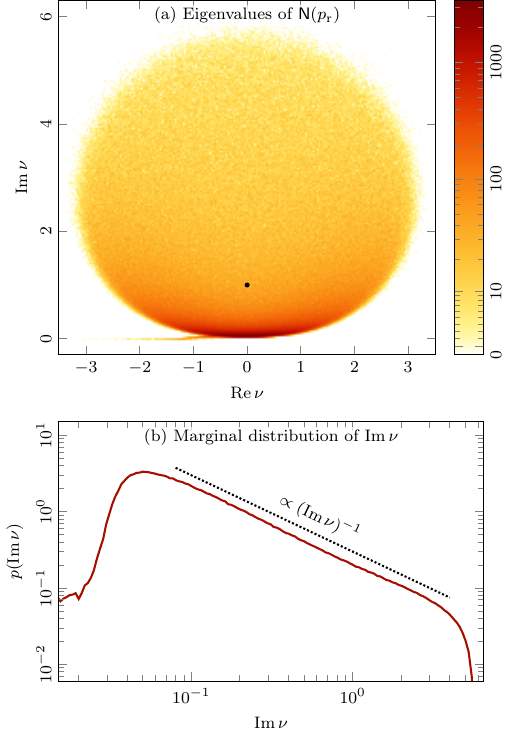}%
\caption{Distribution of the eigenvalues of $\matr{N}(\freal{p})$ in a 2D disk-shaped Lorentz gas for $N=1000$ and $\freal{p}=10\,\varsigma^{-1}$, combining $2^{10}$ random configurations.
Panel (a): Full distribution in the complex $\nu$ plane. The black dot in the bottom depicts the average position~\eqref{eq:nu-average-position}.
Panel (b): Corresponding marginal distribution $p(\Im\nu)$.}
\label{fig:mu-plane-d2-n1000}%
\end{figure}%
\par In the limit of large system ($N\rightarrow\infty$), the eigenvalues distribution of $\matr{N}(\freal{p})$ is shown in Fig.~\ref{fig:mu-plane-d2-n1000} for $d=2$.
Similar results can be found in higher dimensions, but this requires more scatterers.
Here, the situation is very different compared to Fig.~\ref{fig:mu-plane-d3-n100}.
In particular, the circular cluster has become so large that it has absorbed the $s$-wave spiral arm ($\nu_+$).
The cluster has also lost its nearly uniform density.
As explained in Ref.~\cite{Goetschy2011a}, this is due to the constraint on the eigenvalues that $\Im\nu>0$ from Eq.~\eqref{eq:nu-positive-imag}.
In some way, the eigenvalues feel the presence of the boundary $\Im\nu=0$ and accumulate in this vicinity.
The marginal distribution of the imaginary parts of the eigenvalues is shown in Fig.~\ref{fig:mu-plane-d2-n1000}(b).
It reveals that the eigenvalue density inside the circular cluster behaves as
\begin{equation}\label{eq:imag-nu-empirical-law}
p(\Im\nu) \propto \frac{1}{\Im\nu}  \:.
\end{equation}
Due to Eq.~\eqref{eq:resonance-pole-approx}, the same behavior is expected for the resonance density in the vicinity of the single-scatterer resonance $k=p$.
The power law~\eqref{eq:imag-nu-empirical-law} is not obvious, because, at first sight, it reminds us of the Marchenko-Pastur distribution~\eqref{eq:marchenko-pastur-law}.
However, it turns out that Eq.~\eqref{eq:marchenko-pastur-law} behaves as $x^{-1/2}$ for large variance ($\lambda_-\rightarrow 0$), and this is not compatible with the behavior $x^{-1}$ of Eq.~\eqref{eq:imag-nu-empirical-law}.
\par Distributions going as Eq.~\eqref{eq:imag-nu-empirical-law} were also observed numerically in Ref.~\cite{Pinheiro2004} and later explained analytically in Refs.~\cite{Skipetrov2011, Goetschy2011a, Goetschy2013-arxiv}.
To calculate the eigenvalue density, these authors have used a self-consistent equation for the resolvent based on the Cholesky-type decomposition of the Green matrix introduced in Ref.~\cite{Goetschy2011a}.
Here, we numerically confirm that the behavior~\eqref{eq:imag-nu-empirical-law} does not depend on the number of spatial dimensions or the shape of the Lorentz gas.
This supports the idea that this behavior is universal for sufficiently large $N$~\cite{Pinheiro2004, Goetschy2013-arxiv}.

\subsection{Resonance distribution}\label{sec:resonance-density-resonant-model}
Although the eigenvalue approach leads to simple expressions for the resonances, it suffers from two drawbacks.
First, this method finds only at most $N$ resonances.
It does not give any hint about the other resonances located outside the cluster shown in Figs.~\ref{fig:mu-plane-d3-n100} and~\ref{fig:mu-plane-d2-n1000}.
A study of the resonance distribution following this approach is thus limited to a small region of the complex $k$ plane of the order of $L^{-1}$.
Second, this method assumes that the single-scatterer scattering amplitude displays a narrow isolated resonance which may be too restrictive for some applications.
The hard-sphere $s$-wave model~\eqref{eq:point-ampli-hard-sphere}, in particular, has no resonance for $k\in\mathbb{C}$.
Thus, the eigenvalue approach will not help us to find all the complex resonances of the problem, even for the resonant model~\eqref{eq:point-ampli-resonant}.
More generally, we need to resort to the resonance potential method presented in Sec.~\ref{sec:potential-function}.
\par
\begin{figure}[ht]%
\includegraphics{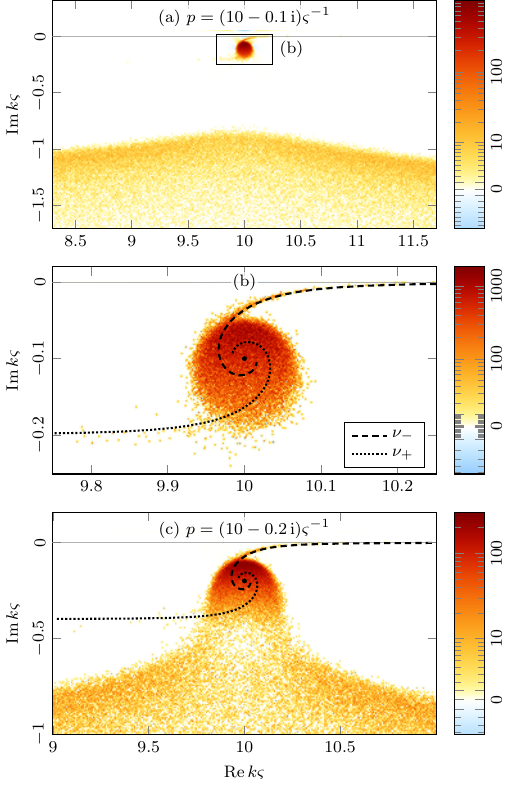}%
\caption{Resonance density $2\pi\rho^{(2)}(k)$ in the complex $k$ plane for the resonant model~\eqref{eq:point-ampli-resonant} in a 3D ball-shaped Lorentz gas with $N=100$.
All panels are numerically computed from Eq.~\eqref{eq:resonance-density-from-potential} with $2^{8}$ random configurations of the scatterers.
Panel (a): Distribution for $p=(10-0.1\,\I)\varsigma^{-1}$ with details shown in panel~(b).
Panel (c): Distribution for the wider resonance $p=(10-0.2\,\I)\varsigma^{-1}$.
The spirals are given by Eqs.~\eqref{eq:resonance-pole-approx} and~\eqref{eq:nu-eigenvalues-two-atoms}.}%
\label{fig:k-plane-resonant}%
\end{figure}

 Full resonance distributions numerically obtained by the resonance potential method are shown in Fig.~\ref{fig:k-plane-resonant}.
These distributions are based on the model of resonant scatterers from Eq.~\eqref{eq:point-ampli-resonant} setting the resonance pole at $p=(10-0.1\,\I)\varsigma^{-1}$ for panels~(a)--(b) and $p=(10-0.2\,\I)\varsigma^{-1}$ for panel~(c).
In panels~(a)--(b) of Fig.~\ref{fig:k-plane-resonant}, one can see an almost circular cluster of resonances centered on the position of the single-scatterer resonance ($k=p$).
The circular cluster in panel~(b) bears some resemblance to the eigenvalue distribution in Fig.~\ref{fig:mu-plane-d3-n100}(a) obtained with the same parameters in combination with the value $\fimag{p}=-0.1\,\varsigma^{-1}$.
The point reflection between Figs.~\ref{fig:mu-plane-d3-n100}(a) and~\ref{fig:k-plane-resonant}(b) is due to the negative value of $\fimag{p}$ in Eq.~\eqref{eq:resonance-pole-approx}.
\par The spiral curves in panels~(b)--(c) of Fig.~\ref{fig:k-plane-resonant} are given by Eqs.~\eqref{eq:resonance-pole-approx} and~\eqref{eq:nu-eigenvalues-two-atoms}.
As explained previously, these resonances result from the proximity between pairs of scatterers, hence the name of proximity resonances~\cite{HellerEJ1996, Rusek2000, Li2003}.
In particular, one notices that the $p$-wave resonances, given by $\nu_-$, have the smallest widths and the longest lifetime.
These structures are directly revealed in the complex $k$ plane instead of indirectly from the eigenvalue method~\eqref{eq:resonance-pole-approx}. %
\par Furthermore, the resonance potential method highlights additional structures of larger widths, such as the resonance band in Fig.~\ref{fig:k-plane-resonant}(a).
This band cannot be predicted by the eigenvalues of $\matr{N}(k)$ without considering complex values of $k$.
Although being of lower density than the main cluster, this band seems to affect the resonance distribution in the main cluster.
This alteration is already visible in the vertical elongation of the cluster in Fig.~\ref{fig:k-plane-resonant}(b).
The interference effect between the band and the circular cluster is quite clear in panel~(c) for which the single-scatterer resonance width is larger.
Another effect is the evanescence of the density in the main cluster as the imaginary part increases in absolute value.
This evanescence is more significant in panel~(c) than in panel~(b).
None of these interference effects can be explained by the eigenvalue distribution of Fig.~\ref{fig:mu-plane-d3-n100}.
This shows the limitation of the eigenvalue method.
\par The existence of the resonance band below the cluster in Fig.~\ref{fig:k-plane-resonant} was expected from our calculation of the two-scatterer case in Sec.~\ref{sec:two-atoms}.
In fact, this structure is due to the multiple scattering of the particle in the medium, and not the presence of the single-scatterer resonance.
The position of this band is thus directly related to the mean escape rate $\Gamma_{\rm esc}$ from the Lorentz gas in the absence of single-scatterer resonance.
If we require the resonance cluster deriving from the single-scatterer resonance to be much smaller than the imaginary coordinate of the resonance band so as to prevent interference effects between these structures, we obtain a supplementary validity criterion for the eigenvalue method of Sec.~\ref{sec:multi-eigenvalues}, namely
\begin{equation}\label{eq:eigval-method-validity}
\abs{\fimag{p}}\rho \ll \frac{\Gamma_{\rm esc}}{2v(\freal{p})}  \:,
\end{equation}
where $\rho$ is the cluster radius estimate of Eq.~\eqref{eq:nu-radius}, and $v(k)$ is the group velocity.
This criterion is quite restrictive since, in the limit of large system ($N\rightarrow\infty$), the $\rho$ tends to infinity and the escape rate to zero.

\section{Hard-sphere \texorpdfstring{$s$}{s}-wave scatterers}\label{sec:hard-sphere-model}
The resonance potential method offers the possibility of studying the resonance distribution in the absence of single-scatterer resonances.
It also allows to consider a wider range of complex wavenumbers than the vicinity of the single-scatterer resonance.
In this section, we study the full distribution of the multiple scattering resonances in a large area of the complex plane of $k$.
As already stressed above, the resonant model~\eqref{eq:point-ampli-resonant} is not physically meaningful far away from the resonance pole $k=p$.
This is why, in this section, we use the more general hard-sphere $s$-wave model~\eqref{eq:point-ampli-resonant} which is valid for $\abs{k}\ll\alpha^{-1}$.
To ensure this condition, we will set the scattering length to $\alpha=0.1\,\varsigma$ in all the figures.
The figures are not significantly affected by this precise choice.

\subsection{Full resonance distribution}\label{sec:resonance-density-hard-sphere}
\begin{figure}[ht]%
\includegraphics{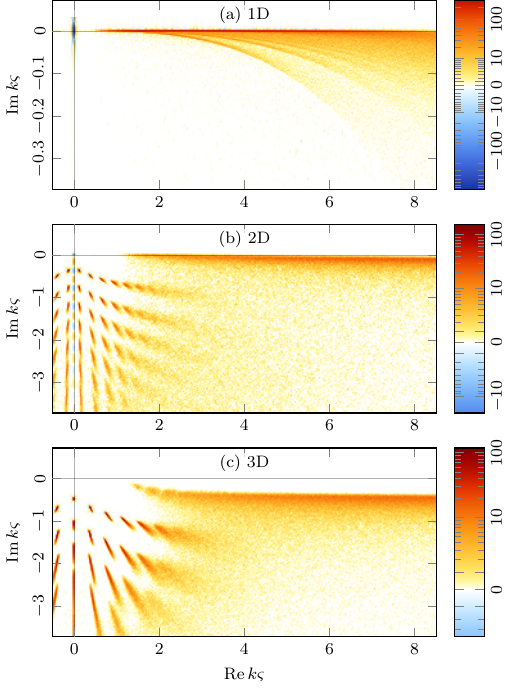}%
\caption{Resonance density $2\pi\rho^{(2)}(k)$ numerically computed from Eq.~\eqref{eq:resonance-density-from-potential} for the non-resonant hard-sphere $s$-wave model~\eqref{eq:point-ampli-hard-sphere} with $\alpha=0.1\,\varsigma$.
Panel~(a): 1D Lorentz gas for $N=10$ and $2^{15}$ configurations.
Panel~(b): 2D Lorentz gas for $N=50$ and $2^{8}$ configurations.
Panel~(c): 3D Lorentz gas for $N=100$ and $2^{8}$ configurations.}%
\label{fig:k-plane-1}%
\end{figure}

Numerical resonance densities obtained from Eq.~\eqref{eq:resonance-density-from-potential} in the three commonest dimensions are shown in Fig.~\ref{fig:k-plane-1}.
First of all, note that the spurious negative densities found in panels~(a) and~(b) are mainly due to the non-analytic singularities of $\det\matr{M}(k)$ and are not meaningful.
These negative densities are located in the regions excluded from the domain $\mathcal{D}$ of Eq.~\eqref{eq:det-m-analytic-domain}, in particular the point $k=0$ in 1D, and the semi-axis $\arg k=-\tfrac{\pi}{2}$ in 2D.
\par Furthermore, the symmetry of imaginary axis in panel~(b) is ensured by our definition~\eqref{eq:free-green-bessel-k} of the Green function.
This symmetry would not have been maintained if we had defined the Green function as Eq.~\eqref{eq:free-green-hankel}, because the branch cut is oriented differently.
\par One striking feature is that panel~(a) of Fig.~\ref{fig:k-plane-1} looks very different from the other panels.
The resonance density in 1D is much more concentrated near the real $k$ axis.
This means that the characteristic escape time is much longer than in higher dimensions.
This is related to the fact that, in 1D, the wave function undergoes strong Anderson localization~\cite{Anderson1958, ShengP2006}, as we will see later again.
In addition, the distribution displays thin stripes having no counterpart in higher dimensions.
\par In panels~(b) and~(c) of Fig.~\ref{fig:k-plane-1}, two different regions are observed: a low-energy region for $\Re k\varsigma\lesssim\pi$ with peaks in the resonance density, and a high-energy region for $\Re k\varsigma\gtrsim\pi$ where the resonances form an almost horizontal band.
At low energy, the distance between two consecutive peaks scales as the inverse of the gas radius.
Preliminary calculations show that these peaks are related to the eigenmodes of the average potential generated by all the scatterers.
Indeed, when the wavelength is larger than the mean inter-scatterer distance, one may expect that the incident wave feels a continuous potential since it cannot resolve the individual scatterers.
\par The resonance band for $\Re k\varsigma\gtrsim\pi$ has the same physical origin as the band seen in Fig.~\ref{fig:k-plane-resonant}(a) and~\ref{fig:k-plane-resonant}(c).
The main difference is that, in Fig.~\ref{fig:k-plane-1}, the resonance distribution is free from single-scatterer resonant clusters.
Only background resonances resulting solely from the multiple scattering remain.
Another difference is that the typical depth of the resonance band with respect to the real $k$ axis is smaller in Fig.~\ref{fig:k-plane-1} than in Fig.~\ref{fig:k-plane-resonant}.
This is due to the larger cross section of the hard-sphere $s$-wave model compared to the resonant model, as shown in Fig.~\ref{fig:scattering-models}.
The larger the cross section, the smaller the escape rate, and the closer the resonance band to the real $k$ axis.
\par The situation in higher dimensions ($d\geq 4$) is very similar to panels~(b) and~(c) of Fig.~\ref{fig:k-plane-1}, except that all the resonances move away from the real $k$ axis, as a consequence of the dimensional reduction of the gas size for fixed $N$.
A detailed study of all the structures of Fig.~\ref{fig:k-plane-1} is deferred to future works.

\subsection{Distribution of the resonance widths}\label{sec:resonance-width}
One particularly interesting feature is the behavior of the resonance density in the resonance band ($\Re k\varsigma\gtrsim\pi$), especially for large negative imaginary parts.
In this region, the resonance potential as well as the density no longer depend so much on $\Re k$ but mainly on $\Im k$.
As one can see in Fig.~\ref{fig:k-plane-1}, the density smoothly decreases when $\Im k\rightarrow-\infty$.
It seems that this decreasing behavior can be roughly approached by the upper bound on the resonance potential $\Phi(k)$ that we derive in this subsection.
\par First, we know that the determinant of any positive-definite matrix can be bounded by its trace. So, we can write
\begin{equation}\label{eq:determinant-upper-bound}
\det(\herm{\matr{M}}\matr{M})^{\frac{1}{N}} \leq \frac{1}{N}\Tr(\herm{\matr{M}}\matr{M})  \:.
\end{equation}
In terms of the eigenvalues of $\herm{\matr{M}}\matr{M}$, this is a consequence of the fact that the geometric mean is always smaller than the corresponding arithmetic mean~\cite{Olver2010}.
Inequality~\eqref{eq:determinant-upper-bound} leads to the following upper bound on the resonance potential~\eqref{eq:def-resonance-potential}
\begin{equation}\label{eq:potential-upper-bound-1}
\Phi(k) = \frac{1}{2N}\avg{\ln\det(\herm{\matr{M}}\matr{M})} \leq \frac{1}{2}\avg{\ln\!\left(\frac{1}{N}\Tr(\herm{\matr{M}}\matr{M})\right)}  \:.
\end{equation}
In addition, we also know that $\avg{\ln X}\leq\ln\avg{X}$ for any positive-definite random variable $X$.
This is again due to the inequality of arithmetic and geometric means~\cite{Olver2010}.
Therefore, a slightly weaker bound than Eq.~\eqref{eq:potential-upper-bound-1} is
\begin{equation}\label{eq:potential-upper-bound-2}
\Phi(k) \leq \frac{1}{2}\ln\!\left(\frac{1}{N}\avg{\Tr(\herm{\matr{M}}\matr{M})}\right)  \:.
\end{equation}
Using Eq.~\eqref{eq:def-m-matrix}, the trace in Eq.~\eqref{eq:potential-upper-bound-2} can be expressed in terms of the matrix elements as
\begin{equation}\label{eq:avg-trace-m2}
\avg{\Tr(\herm{\matr{M}}\matr{M})} = N\abs{F(k)^{-1}}^2 + N(N-1)\avg{\abs{G^+(k,s)}^2}  \:.
\end{equation}
According to Eqs.~\eqref{eq:point-ampli-hard-sphere} and~\eqref{eq:def-free-green-imag}, the asymptotic behavior of $F(k)^{-1}$ for large negative $\Im k$ reads
\begin{equation}\label{eq:invf-deep-k-behavior}
F(k)^{-1} \xrightarrow{\fimag{k}\rightarrow -\infty} 2\I I(k,0)  \:.
\end{equation}
Therefore, $\abs{F(k)^{-1}}^2$ behaves as the power law $\abs{k}^{2(d-2)}$, and is negligible in front of the second term in Eq.~\eqref{eq:avg-trace-m2} which increases exponentially for $\fimag{k}\rightarrow-\infty$.
Indeed, as shown in the~\hyperref[app:green-average]{Appendix}, this term asymptotically behaves as
\begin{equation}\label{eq:avg-g2-complex}
\avg{\abs{G^+(k,s)}^2} \xrightarrow{\fimag{k}\rightarrow-\infty} \frac{d^2\abs{k}^{d-3}}{(2\pi)^{\frac{d-1}{2}} S_dR^{d-1}} \frac{\E^{-4\fimag{k}R}}{(-4\fimag{k}R)^{\frac{d+3}{2}}}  \:.
\end{equation}
Therefore, in the limit $\fimag{k}\rightarrow-\infty$, the upper bound~\eqref{eq:potential-upper-bound-2} varies as
\begin{equation}\label{eq:potential-upper-bound-approx}
\Phi(k) \leq \frac{d-3}{2}\ln\abs{k} - 2\fimag{k}R - \frac{d+3}{4}\ln(-\fimag{k}R) + \mathrm{cst}  \:.
\end{equation}
The result~\eqref{eq:potential-upper-bound-approx} means that the resonance potential does not increase faster than linearly with $-\fimag{k}$ as $\fimag{k}\rightarrow-\infty$.
This constrains the behavior of the resonance density to power laws
\begin{equation}\label{eq:density-upper-bound}
\rho^{(2)}(k) \xrightarrow{\fimag{k}\rightarrow -\infty} (-\fimag{k})^{-\beta}  \qquad\forall\beta>1  \:.
\end{equation}
Indeed, the limit case $\beta=1$ corresponds to the potential $\Phi(k)\propto\bigo(-\fimag{k}L\ln(-\fimag{k}L))$, which is superlinear and exceeds the bound~\eqref{eq:potential-upper-bound-approx}.
However, Eq.~\eqref{eq:density-upper-bound} does not exclude a possible exact cancellation of the density beyond a certain value of $-\fimag{k}$.
This would correspond to $\beta\rightarrow +\infty$ which is allowed by $\beta>1$.
\par
\begin{figure}[ht]%
\includegraphics{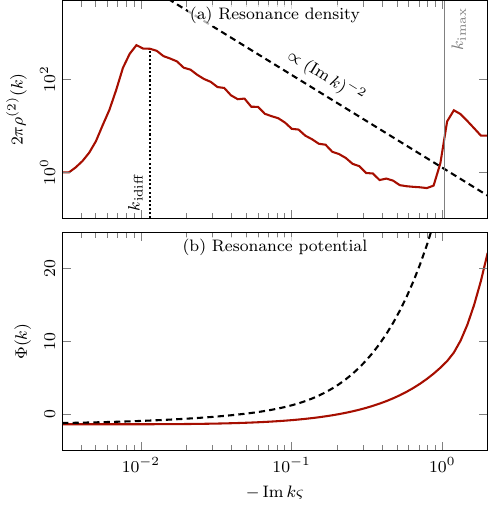}%
\caption{Cross-sectional view of the complex $k$ plane for a 2D Lorentz gas with $N=1000$ and $\alpha=0.1\,\varsigma$ along the vertical axis $\freal{k}=10\,\varsigma^{-1}$.
The curves are averaged over $2^{8}$ configurations. The scale parameter is $n\sigma R\simeq 7.04$.
Panel (a): Resonance density given by Eq.~\eqref{eq:resonance-density-from-potential} (solid), and the approximation~\eqref{eq:density-rough-approx} (dashed). The vertical dotted line is the diffusion approximation~\eqref{eq:diffusion-rate}. The rightmost peak is a numerical artifact.
Panel (b): Corresponding resonance potential (solid), and the upper bound~\eqref{eq:potential-upper-bound-2} (dashed).}%
\label{fig:k-plane-cut-d2}%
\end{figure}

 When the upper bound~\eqref{eq:potential-upper-bound-approx} is used as an approximation of $\Phi(k)$ and inserted into Eq.~\eqref{eq:resonance-density-from-potential}, the power law
\begin{equation}\label{eq:density-rough-approx}
\left(\pder[2]{}{\freal{k}} + \pder[2]{}{\fimag{k}}\right)\Phi(k) \approx \frac{d+3}{4\fimag{k}^2}  \:,
\end{equation}
is obtained for the distribution of the resonance widths at $\Re k\varsigma\gtrsim\pi$.
The approximation~\eqref{eq:density-rough-approx} is compared to the numerical resonance density for a 2D Lorentz gas in Fig.~\ref{fig:k-plane-cut-d2}(a).
This figure depicts the cross-sectional view of the density and the potential along the vertical axis $\freal{k}=10\,\varsigma^{-1}$.
We have chosen this value as a compromise to avoid, on one hand, the distortion of the density due to the low-energy structures in Fig.~\ref{fig:k-plane-1} and, on the other hand, the continuous decrease of the single-scatterer cross section with $\freal{k}$.
\par
\begin{figure}[ht]%
\includegraphics{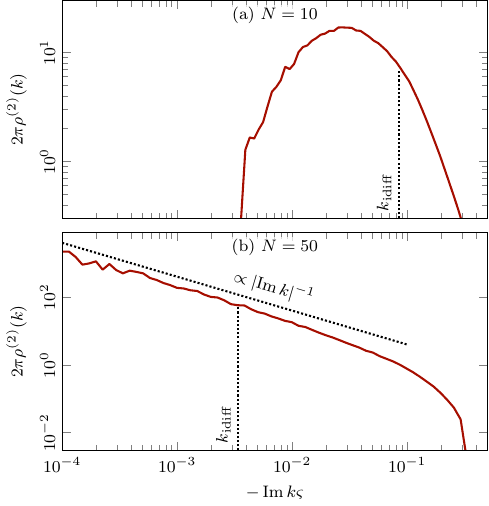}%
\caption{Cross-sectional view of resonance densities for a 1D Lorentz gas with $\alpha=0.1\,\varsigma$ and $\freal{k}=10\,\varsigma^{-1}$. The vertical dotted line is the diffusion approximation~\eqref{eq:diffusion-rate}.
Panel (a): Density for $N=10$ averaged over $2^{22}$ configurations. Panel (b): Density for $N=50$ averaged over $2^{19}$ configurations.}%
\label{fig:k-plane-cut-d1}%
\end{figure}

 It should be noted that our method suffers from significant numerical round-off errors when the imaginary part of $k$ is too negative.
This is also a consequence of the exponential increase of the matrix elements of $\matr{M}(k)$.
One may expect that the method is valid as long as the condition number of $\matr{M}(k)$, denoted as $\kappa$, is smaller than $1/\numeps$, $\numeps\simeq 10^{-16}$ being the machine epsilon in double precision~\cite{Olver2010}.
The condition number of $\matr{M}(k)$ can be roughly estimated as the ratio between the largest and the smallest matrix elements
\begin{equation}\label{eq:condition-number}
\kappa \approx \frac{\abs{G^+(k,L)}}{\abs{G^+(k,\varsigma)}} \approx \E^{-\fimag{k}L}  \:,
\end{equation}
where $L=2R$ is the largest distance between two scatterers.
The range of validity of our method is thus
\begin{equation}\label{eq:numerical-validity}
\fimag{k} > -k_{\rm imax} = \frac{\ln\numeps}{L}  \:.
\end{equation}
The value of $k_{\rm imax}$ is highlighted by the vertical gray line in Fig.~\ref{fig:k-plane-cut-d2}.
Given the restriction~\eqref{eq:numerical-validity}, the existence of the rightmost peak in Fig.~\ref{fig:k-plane-cut-d2} cannot be confirmed.
This structure is probably a numerical artifact and should be ignored.
In spite of this, the plotting regions in Fig.~\ref{fig:k-plane-cut-d1} are fully contained in the validity domain~\eqref{eq:numerical-validity}.
So, the curves are not too much affected by round-off errors.
\par One remarkable point in Fig.~\ref{fig:k-plane-cut-d2} is that the density approximately decreases as $\fimag{k}^{-2}$.
This behavior is quite different from the power law $\abs{\fimag{k}}^{-1}$ obtained in Fig.~\ref{fig:mu-plane-d2-n1000} for the cluster of single-scatterer resonances.
Indeed, the resonances of the horizontal band originate in the many collisions themselves, and not in the existence of a single-scatterer resonance as in Fig.~\ref{fig:k-plane-resonant}.
\par The behavior as $\fimag{k}^{-2}$ seems to be known for weakly open systems in the diffusive regime~\cite{Kottos2005}.
It suggests that the resonances of the random Lorentz gas model could be approached by an effective non-Hermitian Hamiltonian.
However, it is not possible to cast the nonlinear determinantal equation~\eqref{eq:determinantal-equation} into a linear eigenvalue problem for all $k\in\mathbb{C}$.
Therefore, the connection between the behavior $\fimag{k}^{-2}$ and the literature~\cite{Fyodorov1997a, Fyodorov2015, Kottos2005, Weiss2006} is not obvious.
\par Besides of this, it turns out that the abscissa of the left peak in Fig.~\ref{fig:k-plane-cut-d2}(a) approximately corresponds to the escape rate associated with classical diffusion~\cite{Akkermans2007, ShengP2006}, that is to say
\begin{equation}\label{eq:diffusion-rate}
k_{\rm idiff} = -\frac{\ell}{2d}\left(\frac{j_{\frac{d-2}{2}}}{R}\right)^2  \:,
\end{equation}
where $\ell=(n\sigma_{\rm pt})^{-1}$ is the scattering mean free path, and $j_{\frac{d-2}{2}}$ denotes the first zero of the Bessel function $J_{\frac{d-2}{2}}(z)$.
This means that the propagation of the quantum particle is mostly diffusive, and little affected by Anderson localization.
This results from the large value of $k\ell\simeq 25$ compared to 1 given the scattering mean free path $\ell\simeq 2.5\,\varsigma$.
In this regime, localization is not expected to be significant~\cite{Akkermans2007, ShengP2006}.
However, Anderson localization could nevertheless be observed in 2D when $k\lesssim\ell^{-1}$.
\par Finally, resonance densities in a 1D Lorentz gas are shown in Fig.~\ref{fig:k-plane-cut-d1} for two different numbers of scatterers.
In this case, the approximation~\eqref{eq:density-rough-approx} largely overestimates the actual density.
In contrast to the other dimensions, the overall look of the distribution is greatly influenced by the number of scatterers.
When $N$ is small enough, as in panel~(a) of Fig.~\ref{fig:k-plane-cut-d1}, the density displays a peak near diffusion rate $k_{\rm idiff}$ given by Eq.~\eqref{eq:diffusion-rate}.
This peak disappears for moderately large $N$, letting the resonances to accumulate themselves on the real $k$ axis with density $\abs{\fimag{k}}^{-1}$, as shown in panel~(b) of Fig.~\ref{fig:k-plane-cut-d1}.
In this case, the escape time tends to infinity and the particle is blocked in the medium.
Accordingly, we interpret this accumulation of resonances as a consequence of Anderson localization.
Indeed, Anderson localization is expected to take place in one-dimensional disordered systems~\cite{Anderson1958, Anderson1978, ShengP2006, Lagendijk2009}.
In addition, resonance width distributions behaving as $\abs{\fimag{k}}^{-1}$ are known to be a signature of Anderson localization~\cite{Kottos2005}.

\section{Conclusions}\label{sec:conclusions}
We considered the multiple scattering model of a quantum particle in a random Lorentz gas made of point scatterers.
In particular, we studied the distribution of scattering resonances in the complex plane of the wavenumber ($k\in\mathbb{C}$).
To this end, we introduced an efficient method to compute the actual distribution of the resonances in the complex plane of the wavenumber without finding them one by one. We refer to this method as the resonance potential method.
We applied this method for two scattering models for the individual scatterers: a resonant model and a hard-sphere $s$-wave model. The former displays an isolated resonance at a given position $k=\freal{p}+\I\fimag{p}$, in contrast to the latter, which has no resonance. On the other hand, the latter exhibits the expected behavior at low energy.
With the resonant model, two kinds of structures are observed, namely a cluster surrounding the single-scatterer resonance, and a resonance band coming from the multiple collisions themselves.
In addition, the resonance cluster exhibits spiral arms which are interpreted as proximity resonances~\cite{HellerEJ1996, Rusek2000, Li2003}.
We showed that, for small enough $\abs{\fimag{p}}$, the resonance cluster is well described by the eigenvalue distribution of the matrix $\matr{N}(\freal{p})$ defined in Eq.~\eqref{eq:def-normalized-matrix}.
This result confirms the eigenvalue method developed in the literature~\cite{HellerEJ1996, Rusek2000, Pinheiro2004}.
\par Furthermore, the resonance potential method gave us the opportunity to consider the non-resonant hard-sphere model of scatterers.
Indeed, our method does not require the scattering model to be strongly resonant in the region of interest.
With the hard-sphere model, we mapped large regions of the complex $k$ plane hence revealing rich structures that, to our knowledge, were never seen before.
In particular, we observed a collection of peaks at low energy ($\Re k\varsigma\lesssim\pi$), probably coming from a continuum approximation of the Lorentz gas.
At high energy ($\Re k\varsigma\gtrsim\pi$), we highlighted the same kind of resonance band as with the resonant model, but without the distortion due to the resonance cluster.
\par Then, we considered in more details the high energy regime of the hard-sphere model where the resonance density does not significantly depend on the real part of the wavenumber.
We identified the Anderson localization phenomenon in the one-dimensional case as the accumulation of resonances on the real $k$ axis with the density $\abs{\Im k}^{-1}$.
In higher dimensions, the resonance density exhibits a maximum point and decreases faster than $\abs{\Im k}^{-1}$ beyond.
The absence of resonance accumulation on the real $k$ axis in these cases suggests the absence of Anderson localization.
Finally, in future papers, we plan to study the intriguing structures of the resonance density in Fig.~\ref{fig:k-plane-1} to understand their physical origin.

\begin{acknowledgments}%
D.G. holds a Research Fellow (ASP - Aspirant) fellowship from the Belgian National Fund for Scientific Research (F.R.S.-FNRS).
This work was also supported by the F.R.S.-FNRS as part of the Institut Interuniversitaire des Sciences Nucléaires (IISN) under Grant Number 4.45.10.08.
\end{acknowledgments}%

\appendix* %
\renewcommand\theequation{A\arabic{equation}}%
\section{Average of the Green function in a ball}\label{app:green-average}
This Appendix aims at calculating the average of the square modulus of the Green function over two randomly chosen points uniformly distributed in a ball of arbitrary dimension.
This average can be expressed as
\begin{equation}\label{eq:avg-g2-integral}
\avg{\abs{G^+(k,s)}^2} = \int_0^L \abs{G^+(k,s)}^2 P(s) \D s  \:,
\end{equation}
where $P(s)$ denotes the probability density of the distance between any pair of points in the domain $\mathcal{V}$, and $L$ is the maximum distance between them.
If we assume that the domain $\mathcal{V}$ is a $d$-ball of radius $R$, then $L=2R$ and the distance distribution in Eq.~\eqref{eq:avg-g2-integral} reads~\cite{Tu2002}
\begin{equation}\label{eq:distance-distribution-d-ball}
P(s) = d\frac{s^{d-1}}{R^d} I_{1-\frac{s^2}{4R^2}}(\tfrac{d+1}{2},\tfrac{1}{2})  \:,
\end{equation}
where $I_z(a,b)$ denotes the regularized beta function defined by~\cite{Olver2010}
\begin{equation}\label{eq:def-reg-beta}
I_z(a,b) = \frac{1}{\Beta(a,b)} \int_0^z u^{a-1} (1-u)^{b-1} \D u \:.
\end{equation}
In Eq.~\eqref{eq:def-reg-beta}, the notation $\Beta(a,b)=\Gamma(a)\Gamma(b)/\Gamma(a+b)$ stands for the beta function.
Another useful representation of the distribution $P(s)$ is
\begin{equation}
P(s) = d\frac{s^{d-1}}{R^d}\left(1 - \frac{s}{R\Beta(\frac{d+1}{2},\frac{1}{2})}\hypf{\frac{1}{2}}{\frac{1-d}{2}}{\frac{3}{2}}{\frac{s^2}{4R^2}}\right) \:,
\end{equation}
where $\hypf{a}{b}{c}{z}$ is the hypergeometric function~\cite{Olver2010}.
Note again that the distance distribution~\eqref{eq:distance-distribution-d-ball} corresponds to uniformly distributed points in the ball.
This distribution behaves as $\bigo(s^{d-1})$, at small distance, and vanishes at $s=L$.
Although the integral~\eqref{eq:avg-g2-integral} using distribution~\eqref{eq:distance-distribution-d-ball} is not known in closed form, accurate approximations can be obtained with the aid of the following asymptotic behavior for large wavenumbers
\begin{equation}\label{eq:green-square-asym}
\abs{G^+(k,s)}^2 \xrightarrow{\abs{k}s\rightarrow\infty} \frac{\abs{k}^{d-3}}{4(2\pi s)^{d-1}}\E^{-2\fimag{k}s}  \:,
\end{equation}
with $k=\freal{k}+\I\fimag{k}$.
\par In the main text, we consider two special cases for which Eq.~\eqref{eq:avg-g2-integral} can be evaluated accurately: real wavenumbers and large negative imaginary wavenumbers.
In the first case, which is encountered in Eq.~\eqref{eq:trace-g2-from-green} of the main text, we have $\fimag{k}=0$, and the Green function~\eqref{eq:green-square-asym} simplifies.
Using the new variable $z=s^2/4R^2$, we can write Eq.~\eqref{eq:avg-g2-integral} as
\begin{equation}\label{eq:avg-g2-integral-1}\begin{split}
\avg{\abs{G^+(k,s)}^2} & = \frac{d\abs{k}^{d-3}}{2(2\pi R)^{d-1}} \left[1 - \frac{1}{\Beta(\frac{d+1}{2},\frac{1}{2})} \right.\\
 & \left.\times \int_0^1 \hypf{\frac{1}{2}}{\frac{1-d}{2}}{\frac{3}{2}}{z} \D z \right] \:.
\end{split}\end{equation}
The integral in Eq.~\eqref{eq:avg-g2-integral-1} has the closed form~\cite{Olver2010}
\begin{equation}\label{eq:avg-g2-integral-2}
\int_0^1 \hypf{\frac{1}{2}}{\frac{1-d}{2}}{\frac{3}{2}}{z} \D z = \Beta(\tfrac{d+1}{2},\tfrac{1}{2}) - \tfrac{2}{d+1}  \:.
\end{equation}
Then, inserting Eq.~\eqref{eq:avg-g2-integral-2} in Eq.~\eqref{eq:avg-g2-integral-1} leads to the result
\begin{equation}\label{eq:avg-g2-real}
\avg{\abs{G^+(k,s)}^2} = \frac{d\,\Gamma(\frac{d+2}{2})}{2\sqrt{\pi}\Gamma(\frac{d+3}{2})} \frac{\abs{k}^{d-3}}{(2\pi R)^{d-1}}  \:.
\end{equation}
In the three commonest dimensions, Eq.~\eqref{eq:avg-g2-real} becomes
\begin{equation}\label{eq:avg-g2-real-common}
\avg{\abs{G^+(k,s)}^2} = \begin{cases}
1/(2\abs{k})^2      & \text{for}~d=1 \:,\\
2/(3\pi^2\abs{k}R)  & \text{for}~d=2 \:,\\
9/(8\pi R)^2        & \text{for}~d=3 \:.
\end{cases}\end{equation}
So, we find in particular Eq.~\eqref{eq:avg-g2-real-d3} of the main text.
Note that, since the asymptotic expansion~\eqref{eq:green-square-asym} is exact for $d=1$ and $d=3$, the corresponding expressions in Eq.~\eqref{eq:avg-g2-real-common} are also exact.
\par In the second case considered near Eq.~\eqref{eq:avg-trace-m2} of the main text, the imaginary part of the wavenumber is very negative ($\fimag{k}\rightarrow-\infty$).
In this case, the integral~\eqref{eq:avg-g2-integral} is not known in closed form, even with the approximation~\eqref{eq:green-square-asym}.
So, we have to resort to an additional approximation based on the behavior of $P(s)$ near the maximum distance $s=L$.
Indeed, due to the exponentially increasing behavior of Eq.~\eqref{eq:green-square-asym}, the contribution to the integral~\eqref{eq:avg-g2-integral} mainly comes from regions near the maximum distance $s=L$.
It is thus appropriate to expand $P(s)$ near $s=L$ as follows
\begin{equation}\label{eq:d-ball-distribution-approx}
P(s) = \frac{A}{2R} \left(\frac{s}{2R}\right)^{d-1} \left(1 - \frac{s}{2R}\right)^{\frac{d+1}{2}}  \:,
\end{equation}
where $A$ is a constant prefactor which reads
\begin{equation}\label{eq:d-ball-prefactor}
A = \frac{d\,2^{\frac{3d+1}{2}}}{\frac{d+1}{2}\Beta(\frac{d+1}{2},\frac{1}{2})}  \:.
\end{equation}
Note that the prefactor~\eqref{eq:d-ball-prefactor} comes from the series expansion of Eq.~\eqref{eq:distance-distribution-d-ball} near $s=2R$ but does not ensure the proper normalization of Eq.~\eqref{eq:d-ball-distribution-approx}.
In order to restore the normalization of Eq.~\eqref{eq:d-ball-distribution-approx}, we would have to write $A=1/\Beta(d,\tfrac{d+3}{2})$ and Eq.~\eqref{eq:d-ball-distribution-approx} would then become a beta distribution~\cite{Olver2010}.
Although this detail does not affect very much the results, we will keep expression~\eqref{eq:d-ball-prefactor}.
\par Now, substituting Eqs.~\eqref{eq:d-ball-distribution-approx} and~\eqref{eq:green-square-asym} into Eq.~\eqref{eq:avg-g2-integral} and using the new variable $u=s/2R$ gives us
\begin{equation}\label{eq:avg-g2-integral-3}\begin{split}
\avg{\abs{G^+(k,s)}^2} & = \frac{A\abs{k}^{d-3}}{4(4\pi R)^{d-1}} \int_0^1 \E^{-4\fimag{k}Ru} \left(1 - u\right)^{\frac{d+1}{2}} \D u  \\
 & = \frac{A\abs{k}^{d-3}}{4(4\pi R)^{d-1}} \frac{1}{\frac{d+3}{2}} \hypm{1}{\frac{d+5}{2}}{-4\fimag{k}R}  \:,
\end{split}\end{equation}
where $\hypm{a}{b}{z}$ denotes the confluent hypergeometric function~\cite{Olver2010}.
In the special case encountered in Eq.~\eqref{eq:avg-g2-integral-3}, this function admits the following asymptotic behavior
\begin{equation}\label{eq:hypm-a1-asym}
\hypm{1}{b}{z} \xrightarrow{z\rightarrow+\infty} \frac{\Gamma(b)\E^z}{z^{b-1}} - \sum_{n=1}^\infty \frac{\Gamma(b)}{\Gamma(b-n)z^n}  \:.
\end{equation}
Therefore, Eq.~\eqref{eq:avg-g2-integral-3} asymptotically behaves as
\begin{equation}\label{eq:avg-g2-complex-result}
\avg{\abs{G^+(k,s)}^2} \xrightarrow{\fimag{k}\rightarrow-\infty} \frac{A\abs{k}^{d-3}}{4(4\pi R)^{d-1}} \frac{\Gamma(\frac{d+3}{2})\E^{-4\fimag{k}R}}{(-4\fimag{k}R)^{\frac{d+3}{2}}}  \:,
\end{equation}
for large negative imaginary part of the wavenumber $k$.
Finally, inserting Eq.~\eqref{eq:d-ball-prefactor} in Eq.~\eqref{eq:avg-g2-complex-result} and rearranging the factors, we obtain Eq.~\eqref{eq:avg-g2-complex} of the main text.

\end{document}